# Interest Rates and Inflation

## A Simple Model Predicting Interest Rate Behavior

Michael Coopersmith[*] and Pascal J. Gambardella[*]

## Abstract

This article is an extension of the work of one of us (Coopersmith, 2011) in deriving the relationship between certain interest rates and the inflation rate of a two-component economic system. We use the well-known Fisher relation between the difference of the nominal interest rate and its inflation adjusted value to eliminate the inflation rate and obtain a delay differential equation. We provide computer simulated solutions for this equation over regimes of interest.

This paper could be of interest to three audiences: those in Economics who are interested in interest and inflation; those in Mathematics who are interested in examining a detailed analysis of a delay differential equation, which includes a summary of existing results, simulations, and an exact solution; and, those in Physics who are interested in non-traditional methods of modeling.

---

[*] mhc2q@virginia.edu and pascalgambardella@gmail.com

# Contents

## 1. Introduction

This article derives the relationship between long and short term interest rates and the inflation rate of a two-component economic system (Section 2.1) consisting of a manufacturing company and a bank. Using the Fisher relation between the difference of the nominal interest rate and its inflation adjusted value (i.e., the actual long term interest rate), and some additional conditions, we eliminate the inflation rate and obtain a logistic delay differential equation for the actual (real) long term interest rate (Section 2.2).

The logistic[†] equation is famous in the field of population dynamics. In 1838, Verhulst introduced it to describe the population growth of an organism that is limited by the capacity of its environment to support its population. Hutchinson (1948) introduced a fixed delay in the equation to account for situations where some organisms have pulses in reproduction because they have a lag time (e.g., for hatching periods) before they can reproduce again (Erneux, 2009).

The delay in our two-component economic system arises because changes in the bank's interest rates take time to influence the manufacturer's prices. The introduction of a delay introduces the possibility of oscillatory solutions, which are not possible in the first order differential equation without a delay.

---

[†] Verhulst actually used the term "logistique." It is not clear why Verhulst chose the term "logistique" (logistics in English) to describe his equation. Erneux (2009) suggests Verhulst, who was influenced by Malthus's "Essay on the Principle of Population", wanted to differentiate his term from Malthusian "logarithmique" (logarithm in English).

We discuss properties of the logistic delay differential equation (Section 2.3) and apply them to our interest and inflation model (Section 3.1) using computer simulated solutions over regimes of interest. We then relax one of our initial conditions and derive a delay differential equation, which under linear transformation reveals itself to be another delay logistic equation (Section 3.2). Finally, we discuss our conclusions (Section 4).

In Appendix A, we test the accuracy of our simulation for the delay logistic equation by comparing our simulated equation with known properties of this equation, and also with an exact solution. These results allowed us to trust the accuracy, at least for our purposes, of the simulations.

## 2. Interest Rate Equations

### 2.1. Interest Rate Differences and Inflation

We previously (Coopersmith, 2011) derived the following relationship between certain specified interest rate differences and the inflation rate:

(2-1) $\dfrac{d(\ln y(t))}{dt} = I(t - t_0)$ (Principle of Proportionality),

where $y = i_L - i_S$, $i_L$ is the long term interest rate, $i_S$ is the short term interest rate, and $t_0$ is a time delay. Equation (2-1) also holds when $y = y^{(n)} = i_L^{(n)} - i_S^{(n)}$ or $y = y^{(a)} = i_L^{(a)} - i_S^{(a)}$, where the superscripts "n" and "a" refer to the nominal and actual interest rates.

In this section, we repeat the derivation of equation (2-1) to provide a context for the rest of this paper; and, in the next section we obtain an equation for the interest rates alone by eliminating the inflation rate.

We derive equation (2-1) for a simple two-party model of a manufacturer and a bank. Assume the manufacturer sells x objects per month at price p and cost c. Hence, the manufacturer's income is $x \cdot (p - c)$. Assume the bank borrows money $m$ at a low (short-term) $i_S$ interest rate, lends it at a higher (long-term) interest rate $i_L$, and $y$ is the difference in interest rates as defined earlier.

Assume the bank and manufacturer incomes are proportional (with constant $k_1$), and the cost is proportional to the price (with constant $k_2$). Then: $k_1 \cdot m \cdot y = x \cdot (p - c) = x \cdot (1 - k_2) \cdot p$, and hence, p is proportional to y (with constant $k_3$): $p = k_3 \cdot y$, where $k_3 = (\frac{m}{x}) \cdot (\frac{k_1}{1 - k_2})$.

We define inflation I as:

(2-2) $I = \lim_{t_2 \to t_1} (\frac{p(t_2) - p(t_1)}{(t_2 - t_1) \cdot (p(t_1))}) = \frac{d(\ln(p(t)))}{dt}$ .

Noting that $\ln(p) = \ln(k_3) + \ln(y)$ and assuming changes in the bank's interest rates take time $t_0$ to influence the manufacturer's prices, equation (2-1) follows from equation (2-2).

## 2.2. An Interest Rate Delay Differential Equation

In this section, we obtain an equation for the interest rates alone by eliminating the inflation rate. We also illustrate qualitative features of solutions to this interest rate equation.

If we use Fisher's equation (Fisher, 1930) to estimate the relationship between inflation and the nominal and actual interest rates, we can eliminate the inflation rate from equation (2-1). Fisher's equation, which applies to both the long and short term interest rates, is expressed as:

(2-3) $I = i^{(n)} - i^{(a)}$.

Equation (2-3) is an approximation, which is valid when: $(i^{(a)} + I) \gg i^{(a)} \cdot I$. For example, if both interest rates are 0.1 (ten percent) the inequality is $20 \gg 1$. Equation (2.3) is good enough for our present purposes.

This equation also leads to: $y^{(n)} = y^{(a)}$, which is consistent with our initial condition that equation (2-1) holds for both. Inserting $y^{(a)}$ and $I = i_L^{(n)} - i_L^{(a)}$ into equation (2-1) yields:

(2-4) $\dfrac{d(\ln(i_L^{(a)}(t) - i_S^{(a)}(t)))}{dt} = i_L^{(n)}(t - t_0) - i_L^{(a)}(t - t_0)$.

Although equation (2-4) appears to connect (and therefore determine) the interest rate as a function of time it actually provides only a starting point because we now have four different interest rates. If we place the following conditions on these interest rates, we can reduce the equation to one involving a single interest rate:

(i) **The long term nominal interest rate is a constant and greater than zero.** An example is the US prime bank lending rate[‡], which is the average rate of interest charged on short term loans by commercial banks to companies. It has been a constant 3.25% per year from January 2009 until November 2015. This condition leads to (2-5) $i_L^{(n)}(t-t_0) = A$, where $A$ is a constant greater than zero.

---

[‡] See http://www.economagic.com/em-cgi/data.exe/fedbog/prime#Monthly.

(ii) The **short term actual interest rate is zero.** An example is the federal funds rate, which has been[§] between 0 and 0.25% from December 2009 through November 2015. This condition leads to:

(2-6) $i_S^{(a)}(t) = 0$.

We will relax this condition in Section 3-2 and define $i_S^{(a)}(t)$ as a positive, non-zero constant.

Using equations (2-5) and (2-6) in equation (2-4) yields: $\dfrac{d(\ln(i_L^{(a)}(t)))}{dt} = A - i_L^{(a)}(t-t_0)$, which upon expansion becomes:

(2-7) $\dfrac{d(i_L^{(a)}(t))}{dt} = A \cdot i_L^{(a)}(t) - i_L^{(a)}(t-t_0) \cdot i_L^{(a)}(t)$.

As mentioned earlier, equation (2-7) has been traditionally called the delay logistic equation or Hutchinson's equation (Erneux 2009, Smith 2011). It is also a functional differential equation where its solutions are determined by an initial function $i_L^{(a)}(t) \equiv \Psi(t)$ defined over the interval $0 \leq t \leq t_0$. This contrasts with the solutions of an ordinary differential equation, where its solutions are determined by an initial value at a point. For example, a unique solution of $\dfrac{dx(t)}{dt} = a \cdot x(t)$ is determined by the initial condition: $x(0) = 0$. The introduction of a delay introduces the *possibility* of oscillatory solutions, which are not possible without the delay.

Combining equations (2-3), (2-5) and (2-6) yields:

---

[§] See http://useconomy.about.com/od/monetarypolicy/p/Past_Fed_Funds.htm.

(2-8) $I(t) = A - i_L^{(a)}(t)$.

Equation (2-8) indicates that solutions of equation (2-7) can be transformed to solutions for *I(t)*.

## 2.3. Delay Logistic Equation

This section explores the solutions and properties of the logistic delayed differential equation. In Section 3, we apply these results to our interest rate equation. A reader aware of these results can skip this section.

Solutions of the delay logistic equation have interesting properties that depend on the value of *A*, where the constant, $A = i_L^{(n)}$. To better describe these properties, collected from many references, and apply them to our interest and inflation model, we first need to transform equation (2-7) into a canonical form where the unit of time is 1. Assume

(2-9) $i_L^{(a)}(t') = z(t)/t_0$, where $t_0 \neq 0$,

(2-10) $t' = t \cdot t_0$, and

(2-11) $A = a/t_0$.

Then, equation (2-7) becomes (after dropping the prime from the *t*) the **discrete** delay logistic equation; called "discrete" because the delay is now 1 instead of $t_0$:

(2-12) $\dfrac{dz(t)}{dt} = -z(t) \cdot z(t-1) + a \cdot z(t)$

We define a solution to equation (2-12) as one which is a continuously differentiable function (i.e., in $C^{(1)}$) for $t \geq 0$. This means the derivative $\dfrac{dz(t)}{dt}$ exists and is itself a continuous function (e.g., it does not have a jump discontinuity). As discussed in the last section, the solutions of equation (2-12) are determined by an initial function. Define $z(t) = \Phi(t)$ within the interval $0 \leq t \leq 1$ as that function. Although there is a wide range of initial functions to choose, we choose those that also satisfy equation (2-12). Kakutani and Markus (1958) have shown that continuously differential solutions of $z(t)$ exist for $t \geq 0$ if and only if $\Phi(t)$ is a continuously differentiable function on $0 \leq t \leq 1$ and

$$(2\text{-}13)\ \frac{d\Phi(1)}{dt} = [a - \Phi(0)] \cdot \Phi(1).$$

The notation $\Phi(c)$ indicates that the quantity is evaluated at the value of $t = c$, where $c$ equals a constant. Here is an example of one function that satisfies equation (2-13):

$$(2\text{-}14)\ \Phi(t) = \beta \cdot e^{(a-\beta)t},\ \text{where } \beta \text{ is a constant greater than zero.}$$

Trivial solutions of equation (2-12) are $z(t) = 0$ and $z(t) = a$. If a non-zero solution becomes zero after a time $t_1$ then $z(t) = 0$ for all $t \geq t_1$ (Cunningham, 1954). This leads to several consequences:

(a) **Solutions can never oscillate about $z(t) = 0$.** However, as we will discuss later, solutions may oscillate about $z(t) = a$.

(b) **A solution $z(t)$ is either entirely positive or negative, but not both.** The sign of the initial function $\Phi(1)$ at $t=1$ controls the sign of $z(t)$.

Qualitative characteristics of the solutions to equation (2-12) (e.g., damped, damped oscillations, sustained oscillations) depend only the value of "$a$" as we will discuss. This means that these characteristics depend on the product of $A$ and $t_0$ in the original equation (see equation 2-7).

The following discussion collects the results for different ranges of "*a"* indicating when the solutions oscillate or not; and, if they oscillate when we have damped or sustaining oscillations. These results are used in Section 3 when we apply what we learn to our interest rate equation.

### 2.3.1. Asymptotic, Non-Oscillatory Solutions: $0 < a \leq (1/e)$

If $0 < a \leq (1/e)$ (where 1/e = 0.367879441…), $\Phi(1) > 0$ and the interval between the zeros of (*z(t)-a*) are at least 1 for large t (e.g., this excludes the $z(t) = 0$ solution), then the solution *z(t)* is asymptotic to *z = a* (Kakutani and Markus,1958, Theorem 10). This means solutions with *a < (1/e)* don't oscillate under these conditions. The condition $\Phi(1) > 0$ implies $z(t) > 0$ since $\Phi(t)$ satisfies equation (2-12), which has solutions which are positive or negative, but not both.

Figure 2.1 illustrates a run[**] using $\Phi(t)$ from equation (2-13) as the initial function with *a=0.35* and $\beta = 0.2$. Notice that the resulting solution is asymptotic to *z(t) = a = 0.35* with initial value $\Phi(0) = 0.2$ as expected.

---

[**] Stella Professional, a system modeling tool (http://www.iseesystems.com/softwares/stella-pro/v1.aspx#), was used to generate solutions.

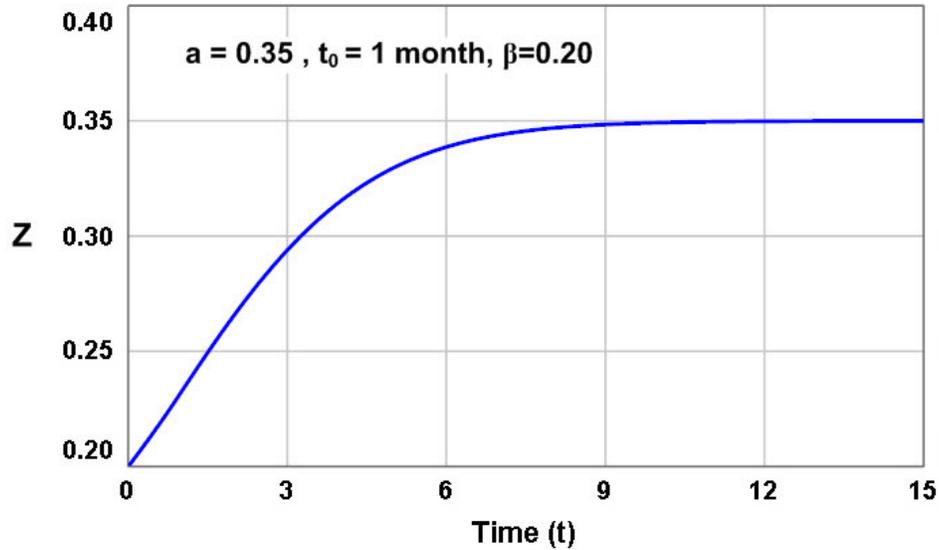

**Figure 2-1.** Example of a Solution Asymptotic to z(t) = a = 0.35

### 2.3.2. Oscillatory Solutions: $(1/e) < a$

Kakutani and Markus (1958, Theorem 6) prove that solutions of equation (2-12) with *a > 0* and large enough t are either (1) asymptotic to *z = a,* or (2) oscillate about *z = a*. Since solutions with "*a*" satisfying $0 < a \leq 1/e$ are purely asymptotic solutions, solutions with "*a*" satisfying $a > (1/e)$ are oscillating solutions. The plot illustrated in Figure 2-2 using $\Phi(t)$ from equation (2-13) as the initial function with *a*=1, and $\beta = 0.12$. Notice that *z(t)* is asymptotic to *a* = 1 as expected.

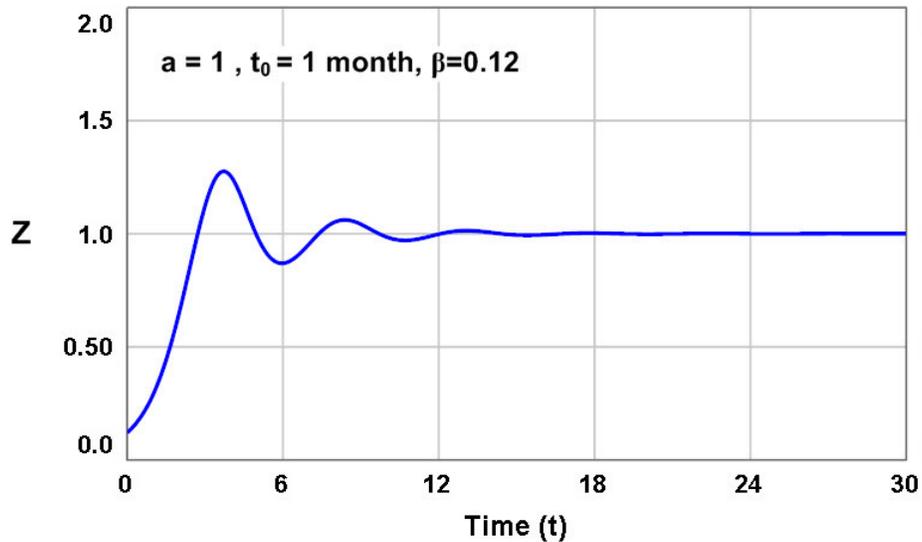
**Figure 2-2**. Example of a Solution with Oscillations that Decay to $z(t) = a = 1$

In the Figure 2.2, the solution $z(t)$ eventually decays into a stable steady-state $z(t) = a$. Does this stable behavior persist for all "$a$"? Figure 2-3 suggests that there is a transition somewhere between $a = 1.5$ and $a = 1.6$ between a solution that decays to a steady state and one that eventually exhibits sustained oscillations. This transition is an example[††] of a Hopf bifurcation (Smith, 2011). It can occur as the birth of a periodic solution from a steady state as a parameter (e.g., like "$a$") crosses a critical value. This critical value of "$a$" is independent of the initial function $\Phi$, as we will discuss next.

---

[††] Touboul (2014) applied methods from Physics to study the "hipster effect", which he defined as "thus non-concerted emergent collective phenomena of looking alike trying to look different." He ended up with a delay differential equation that exhibited a Hopf bifurcation. As in our situation, it also occurred as the birth of a periodic solution from a steady state one as a parameter, the time delay, increased.

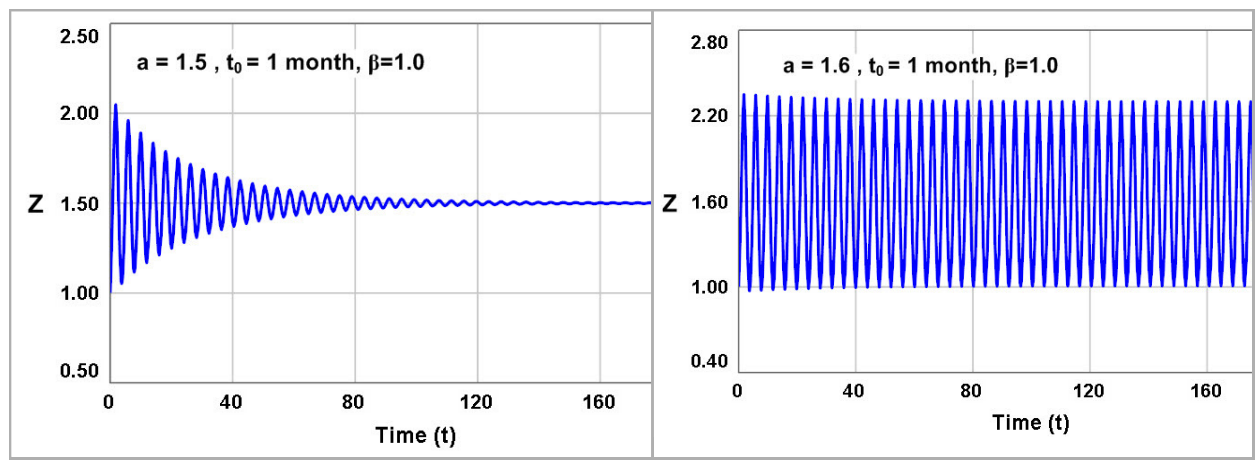

**Figure 2-3**. Solutions illustrating the Change from Steady-State *(a=1.5)* to Non-Steady State *(a = 1.6)*

Kakutani and Markus (1958, Theorems 8) additionally show: if $0 < a < 1$ and $(z(t) - a)$ oscillates with discrete zeros, then the oscillations are damped. This means that for $a < 1$ the solution will be asymptotic for all initial functions $\Phi$ (as defined above) and not transition to sustained oscillations. In this situation, the equilibrium point $z = a$ with $a < 1$ is called globally stable since it does not depend on the initial conditions.

The plot on the left side of Figure 2.3 indicates that the solution is still asymptotic for $a = 1.5$. What is happening here? In an addendum to their paper, Kakutani and Markus said that Wright improved the upper estimate to the bound on "$a$' to 1.5. In addition, Wright conjectured the actual value for this upper limit was $\pi/2$. You can get an upper bound of $\pi/2$ if you look at perturbations of the solution around $z = a$ ((Erneux, 2009). In this situation with an upper bound of $\pi/2$ the equilibrium point $z = a$ is called asymptotically stable because it might depend on the initial conditions.

Is there a proof of global stability for an upper bound of $\pi/2$? Apparently not yet, but some have come close. In a recent paper about Wright's conjecture, Bánhelyi, et al (2014) provided the estimate 1.5706, which is close to $\pi/2 = 1.570796...$ This means we have damped oscillations when $(1/e) < a < 1.5706$.

When a>1.5706 and after a long enough time, the solutions *z(t)* become sustained oscillations (e.g., of equal amplitude) (Bánhelyi, Csendes, Krisztin, and Neumaier, 2014). This is consistent with the results of the simulations illustrated in Figure 2.3 (and in Appendix A1). Many literature references to the results mentioned in this section use the equation:

(2-15) $y(t) = a \cdot y(t) - a \cdot y(t) \cdot y(t-1)$.

This equation (also referred to as Wright's equation) can be transformed to equation (2-12) using the transformation *y = z/a*. Properties characterized by the value of "*a*", as described above, apply to both equations.

What does this all mean for our interest rate model? We discuss this in Section 3.

## 3. Interest Rates and Inflation

### 3.1. Applying Logistic Equation Characteristics

This section applies the results of the Section 2.3 to our model of interest rates and inflation. Define an initial function $\Psi(t)$ as follows:

(3-1) $i_L^{(a)}(t) = \Psi(t) = \beta \cdot e^{(A-\beta)t}$ , where $\beta$ is a constant greater than zero and $0 \leq t \leq t_0$. The initial function $\Psi(t)$ satisfies the equation, which is similar to equation (2-14):

$$\frac{d\Psi(t_0)}{dt} = [A - \Psi(0)] \cdot \Psi(t_0).$$

### 3.1.1. Asymptotic, Non-Oscillatory, Solutions: $0 < i_L^{(n)} \leq \frac{1}{e \cdot t_0}$

These solutions occur when $0 < i_L^{(n)} \leq \frac{1}{e \cdot t_0}$. For large enough t: $i_L^{(a)}(t)$ is asymptotic to $i_L^{(n)} = A$, which means (from equation 3-1) the inflation rate eventually approaches zero. The initial condition in equation (3-1) establishes the initial values of the inflation rate. Assume the interest rates have units of "1/year", and $i_L^{(a)}(t)$ (treated like a time varying widget having its own units) can change monthly, so t and $t_0$ have units of month. Note that

$$I(t) = i_L^{(n)} - i_L^{(a)}(t) = A - i_L^{(a)}(t).$$

If $A = 0.035$ (3.5%/year) then $i_L^{(a)}(t)$ will have asymptotic solutions when $t_0 < 1.05$ months. Figure 3-1 illustrates this when $t_0 = 1$.

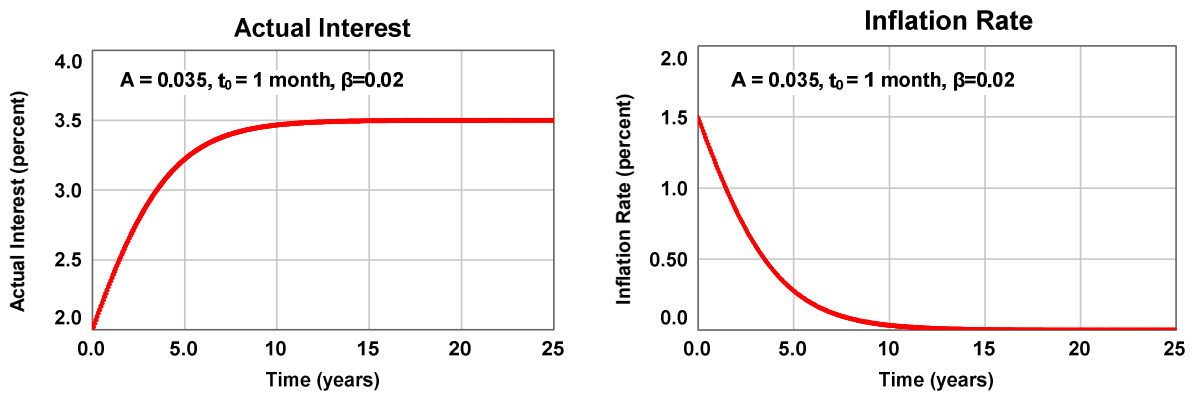

**Figure 3-1.** Example of Asymptotic Behavior in the Actual Interest and Inflation Rates

## 3.1.2. Oscillatory Solutions: $\frac{1}{e \cdot t_0} < i_L^{(n)}$

Oscillatory solutions occur when: $\frac{1}{e \cdot t_0} < i_L^{(n)}$. The oscillations are damped when $i_L^{(n)} < (1.5706/t_0)$. The "actual interest rate" $i_L^{(a)}(t)$ will oscillate about $i_L^{(n)} = A$. If $i_L^{(n)} = 0.08$ (8% per year), then $i_L^{(a)}(t)$ will have damped oscillations when $(1.05 \text{ months}) < t_0 < (44.8 \text{ months})$. Figure 3-2 illustrates this when $t_0 = 12$.

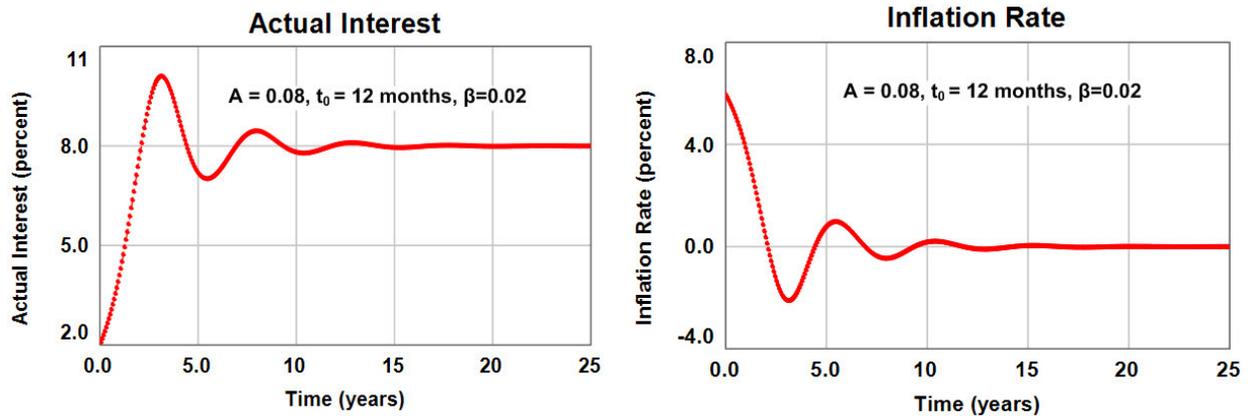

**Figure 3-2.** Example of Damped Oscillations in the Actual Interest and Inflation Rates

In Figure 3-2, the inflation rate oscillates about zero indicating periods of inflation and deflation. Sustained oscillations solutions occur when: $(1.5706/t_0) < i_L^{(n)}$. If $i_L^{(n)} = 0.12$ (12%/year), then $i_L^{(a)}(t)$ will have solutions with sustained oscillations when $t_0 > 13.09$ months, which would indicate a really slow process in communicating interest rates. Figure 3-3 illustrates this situation.

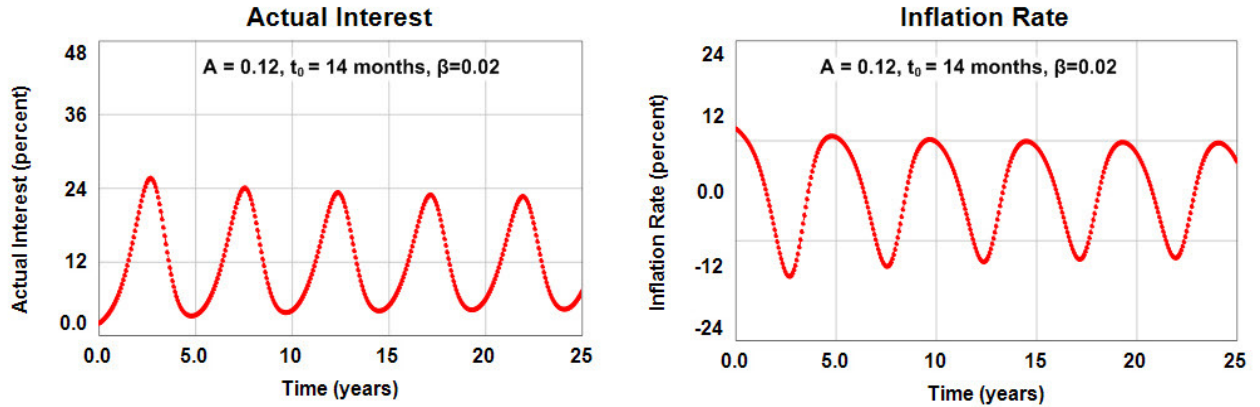

**Figure 3-3.** Example of Sustained Oscillations in the Actual Interest and Inflation Rates

## 3.2. Assuming a Non-zero, Constant Short Term Interest Rate

Let's relax the condition that $i_S^{(a)}(t) = 0$, and instead assume

(3.2) $i_S^{(a)}(t) = w$, where $w$ is a constant, and $w < A$.

Substituting equation (3-2) into equation (2-4) and assuming $i_L^{(n)}(t - t_0) = A$, where $A$ is a constant, yields:

(3-3) $\dfrac{d(i_L^{(a)}(t))}{dt} = A \cdot (i_L^{(a)}(t) - w) - i_L^{(a)}(t - t_0) \cdot (i_L^{(a)}(t) - w)$.

Figure 3-4 shows the simulation results for this equation when $A=0.08$ (8 percent), $w=0$ (delay logistic equation), $w=0.01$ (1 percent), and $w=0.02$ (2 percent). The introduction of a non-zero $w$ increases the inflation rate so that the amount of deflation for each oscillation is less as $w$ increases. Also, when time $t > 243$ months all runs converge to $A = 0.08$. This means, at least for the values of parameters we tested, the introduction of the perturbation $w$ to the delay logistic equation does not change the stable equilibrium.

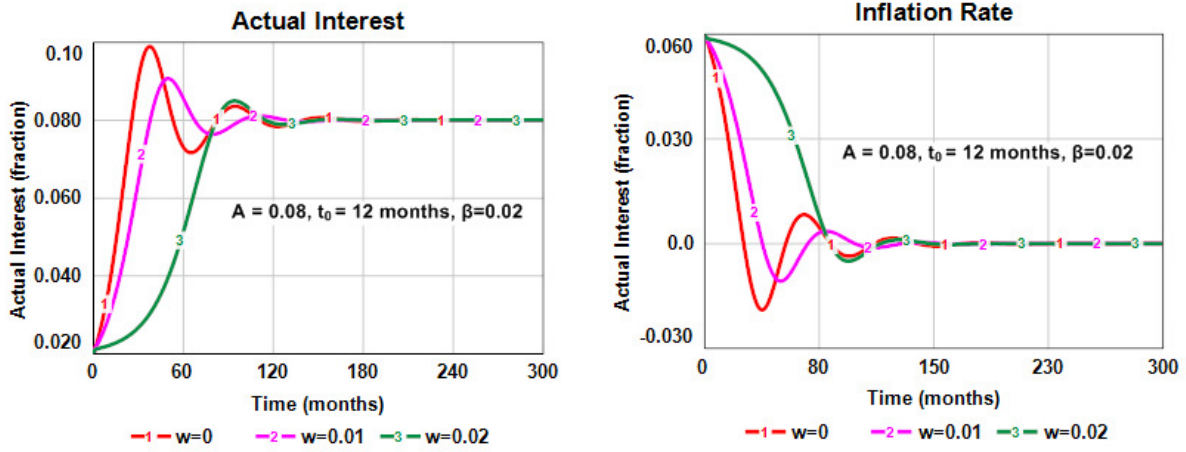

**Figure 3-4.** Example of Simulation of Equation (3-3) with A=0.08 and Various Values of w

Figure 3-5 shows simulated results for this equation when *A*=0.12, *w*=0 (delay logistic equation), *w*=0.01 (1 percent), and *w*=0.02 (2 percent).

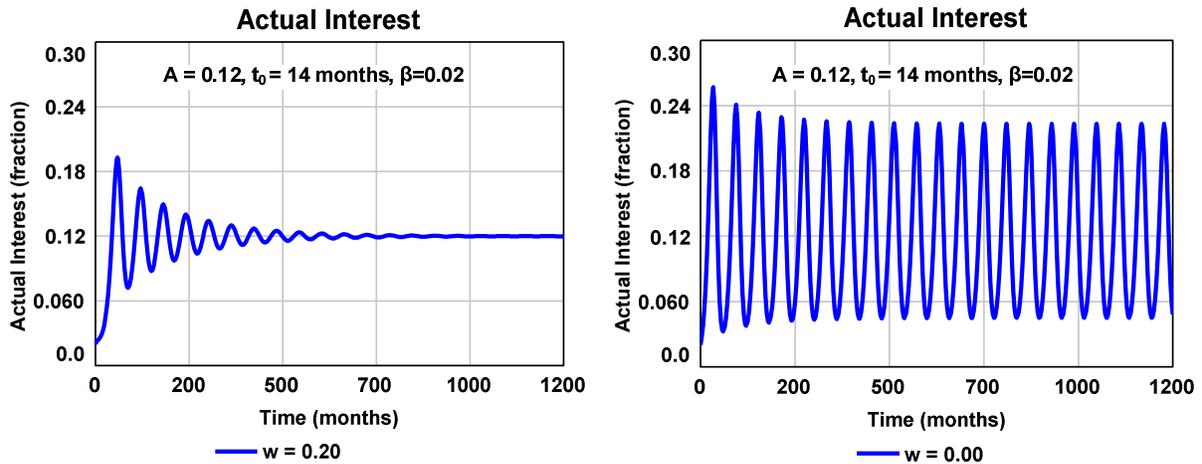

**Figure 3-5.** Example of Simulation of Equation (3-3) with A=0.12 for w=0.20 and w=0.00

What is interesting about the results in Figure 3-5 is that *w* changes from a sustained equilibrium (w=0) to a stable one (w=0.02, $t > 1113$ months) where the actual interest is eventually equal to $A = 0.12$. Having a non-zero short term interest rate can seemly increase the stability of the solution, at least in this example. However, there is more to this result.

After reviewing the results in Figures 3-4 and 3-5, we wondered how closely equation (3-3) relates to a delay logistic equation. It turns out that when we make the transformation $i_L^{(a)}(t) = x(t) + w$ equation (3-3) becomes the delay logistic equation:

$$(3\text{-}4)\ \frac{dx(t)}{dt} = (A - w) \cdot x(t) - x(t) \cdot x(t - t_0).$$

Instead of $a = A \cdot t_0$ being the critical parameter, it is now $a' = (A - w) \cdot t_0$, which explains our results. In figure 3-5, $w = 0$ corresponds to $a' = A \cdot t_0 = 1.68 > 1.507$, thus indicating an evolution to sustained oscillations (see Section 2.3.2); and, $w = 0.2$ corresponds $a' = (A - w) \cdot t_0 = 1.4 < 1.507$, thus indicating an evolution to an asymptotic solution.

To summarize our results:

(3-5) $i_L^{(a)}(t) = x(t) + i_S^{(a)}$, where $i_S^{(a)} (= w)$ is a constant with $i_S^{(a)} \geq 0$,

(3-6) $I(t) = i_L^{(n)} - i_S^{(a)} - x(t)$, where $i_L^{(n)} (= A)$ is a constant with $i_L^{(n)} > i_S^{(a)}$, and

x(t) satisfies the logistic delay differential equation:

$$(3\text{-}7)\ \frac{dx(t)}{dt} = (i_L^{(n)} - i_S^{(a)}) \cdot x(t) - x(t) \cdot x(t - t_0).$$

We will refer to this as the DD (delayed-differential) Interest equation.

Here is an example of an initial condition:

(3-8) $i_L^{(a)}(t) = \Psi(t) = \beta \cdot e^{(A-w-\beta)t}$, where $\beta$ is a constant greater than zero, and $0 \leq t \leq t_0$. The function $\Psi(t)$ satisfied the constraint:

(3-9) $\dfrac{d\Psi(t_0)}{dt} = [A - w - \Psi(0)] \cdot \Psi(t_0)$ .

## 4. Concluding Remarks

Our model, like the real world, can exhibit fluctuations and, concomitantly, periods of inflation and deflation (negative inflation). The cause of this behavior is the communication delay $t_0$ in the model when it is large enough to satisfy the condition (see sections 3.1.2 and 3-2):

(4-1) $(1/e) < (i_L^{(n)} - i_S^{(a)}) \cdot t_0$ .

We interpret the presence of oscillating behavior in our solution as representing chaos in the financial system.

This property of oscillation has implications for monetary policy. For example, since it is desirable that the financial system remain stable and orderly (and avoid oscillations), then any changes instituted by a controlling agency (e.g., the Federal Reserve) should satisfy the inequality:

(4-2) $(1/e) \geq (i_L^{(n)} - i_S^{(a)}) \cdot t_0$ .

A related example is the effect of the bank in our model changing its short term interest rates. If the system is close to the condition for oscillations and the short term actual rate $i_S^{(a)}$ is set to negative while the rate $i_L^{(n)}$ remains the same, then the system may satisfy equation 4-1 and oscillate.

# Appendix A - Test of Simulation Accuracy

The purpose of this appendix is to provide confidence in the use of our simulation application to explore delay differential equations.

## A1. Boundary Tests

One way to test the accuracy of our simulation application is to see how closely it predicts the transition as "$a$" increases from (1) purely asymptotic solutions to damped oscillations (stable state), and from (2) damped oscillations to sustained oscillations (unstable state). In this section we provide results for the second situation.

Figure A-1 illustrates that the transition from damped oscillations (stable state) to sustained oscillations occurs for "$a$" between $a=1.568$ and $a=1.570$. This means the stable state lasts until at least a=1.568. So far it has been proved (see Section 2.3.2) that this stable state lasts until a=15706. Hence, our simulation is at least accurate to three significant figures, which is good enough for our purposes.

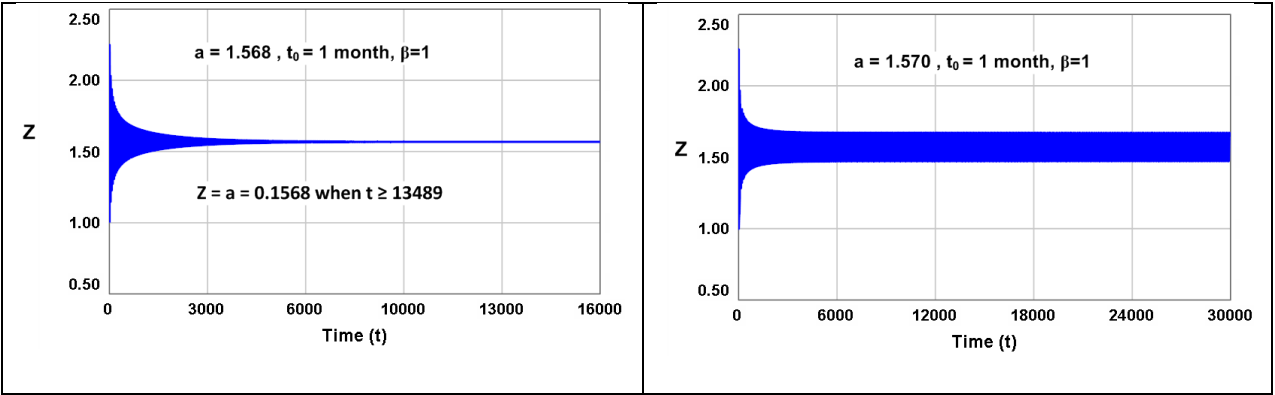

**Figure A-1.** Transition from stable to unstable states

## A2. Exact Solution

Another way to test the accuracy of the solution is to provide an exact solution, and then compare the simulated and exact solutions.

We solved equation (2-12) for "t" within the interval $0 \leq t \leq 3$. We assumed the initial function $\Phi(t)$ in equation (2-13) for the interval $0 \leq t \leq 1$. We then derived $z(t)$ for the two intervals: $1 \leq t \leq 2$ and $2 \leq t \leq 3$. We set $\beta = a/2$ in equation (2-14) to simplify the result. Otherwise, the solution would contain incomplete gamma functions. It would make entering the exact solution in Stella Pro more difficult. Table A-1 contains the solution,

| Interval | Solution | Value at Lower Boundary | Value at Higher Boundary |
|---|---|---|---|
| $0 \leq t \leq 1$ | $z(t) = \Phi(t) = (\frac{a}{2})e^{(a/2)t}$ | $z(0) = \Phi(0) = a/2$ | $z(1) = \Phi(1) = (\frac{a}{2})e^{(a/2)}$ |
| $1 \leq t \leq 2$ | $z(t) = \Phi(1)e^{[a(t-1)-(e^{(a/2)(t-1)}-1)]}$ | $z(1) = \Phi(1)$ | $z(2) = \Phi(1)e^{[a-(e^{(a/2)}-1)]}$ |
| $2 \leq t \leq 3$ | $z(t) = z(2)\exp\{a(t-2) - [e^{\frac{a+2}{2}}(-e^{-e^{(\frac{a}{2})(t-2)}}(e^{(\frac{a}{2})(t-2)}+1) + (\frac{2}{e}))]\}$ | $z(2)$ | $z(3) = z(2)\exp\{a - [e^{\frac{a+2}{2}}(-e^{-e^{(\frac{a}{2})}}(e^{(\frac{a}{2})}+1) + (\frac{2}{e}))]\}$ |

**Table A-1.** Solution of z(t) for $0 \leq t \leq 3$.

Figure A-2 illustrates that, at least graphically, the actual solution matches the simulated solution.

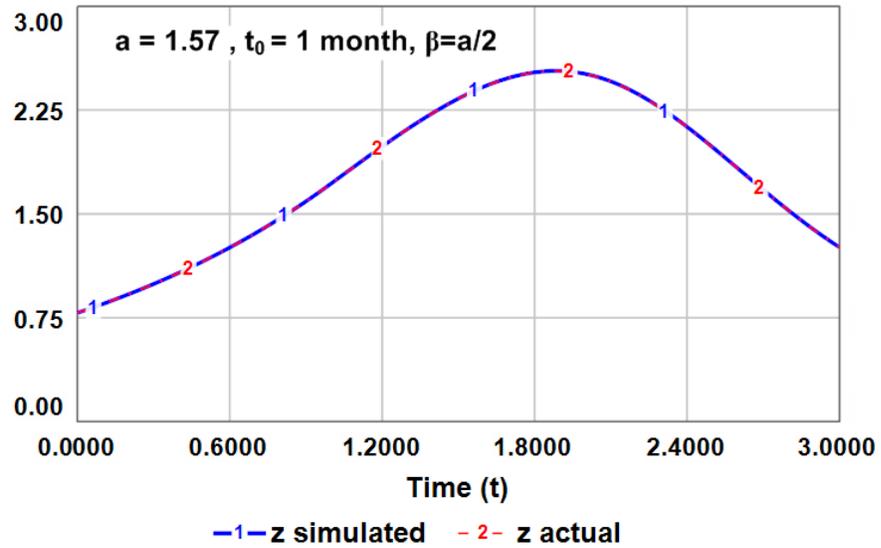

**Figure A-2.** Comparison of Simulated and Actual Solutions

Most of the runs in this paper were done with simulation time step = 1/512 (i.e., step size for the simulation). Table A-1 shows that the errors using this time step are accurate to three significant figures, and if we go to a smaller time step we can improve the accuracy by a factor of 2.

| | Time Step 1/512 | | | Time Step =1/1024 | | |
|---|---|---|---|---|---|---|
| time | z actual | z simulated | delta | z actual | z simulated | delta |
| 0 | 0.785 | 0.785 | 0 | 0.785 | 0.785 | 0 |
| 0.25 | 0.955212394 | 0.955212394 | 0 | 0.955212394 | 0.955212394 | 0 |
| 0.5 | 1.162332124 | 1.162332124 | 0 | 1.162332124 | 1.162332124 | 0 |
| 0.75 | 1.414361848 | 1.414361848 | 0 | 1.414361848 | 1.414361848 | 0 |
| 1 | 1.721039448 | 1.721039448 | 6.35E-14 | 1.721039448 | 1.721039448 | 0 |
| 1.25 | 2.051553608 | 2.051894565 | -0.000340957 | 2.051553608 | 2.051724101 | -0.000170493 |
| 1.5 | 2.333223 | 2.334082704 | -0.000859704 | 2.333223 | 2.333652867 | -0.000429867 |
| 1.75 | 2.506014394 | 2.507554698 | -0.001540304 | 2.506014394 | 2.506784526 | -0.000770132 |
| 2 | 2.510599321 | 2.512894725 | -0.002295404 | 2.510599321 | 2.511746908 | -0.001147586 |
| 2.25 | 2.318651603 | 2.3214308 | -0.002779197 | 2.318651603 | 2.320040707 | -0.001389104 |
| 2.5 | 1.981346675 | 1.98397818 | -0.002631505 | 1.981346675 | 1.982661616 | -0.001314941 |
| 2.75 | 1.597509757 | 1.599427315 | -0.001917558 | 1.597509757 | 1.598467609 | -0.000957851 |
| 3 | 1.258411453 | 1.259325011 | -0.000913558 | 1.258411453 | 1.258867355 | -0.000455903 |

**Table A-2.** Comparing the Actual and Simulated Solutions for two Time Steps.

## Appendix B - Simulation Code

We used Stella Professional from isee systems (http://www.iseesystems.com/softwares/stella-pro/v1.aspx#) to generate all the simulation runs in this paper. We found Stella Professional, a visual, interactive modeling tool, useful in studying delay differential equations. It is primarily used in System Dynamics modeling.

We concur with Jay Forrester, founder of the field of System Dynamics, who suggested that system dynamics (application software) could be useful in solving and understanding differential equations.

We adhere to the policy of the System Dynamics community to include the simulation code used to generate results. Hence, we include a diagram of the code (Figure B-1) taken from Stella Pro, and the code itself. Other System Dynamics software can also be used to generate the results using a diagram and code similar to ones included here.

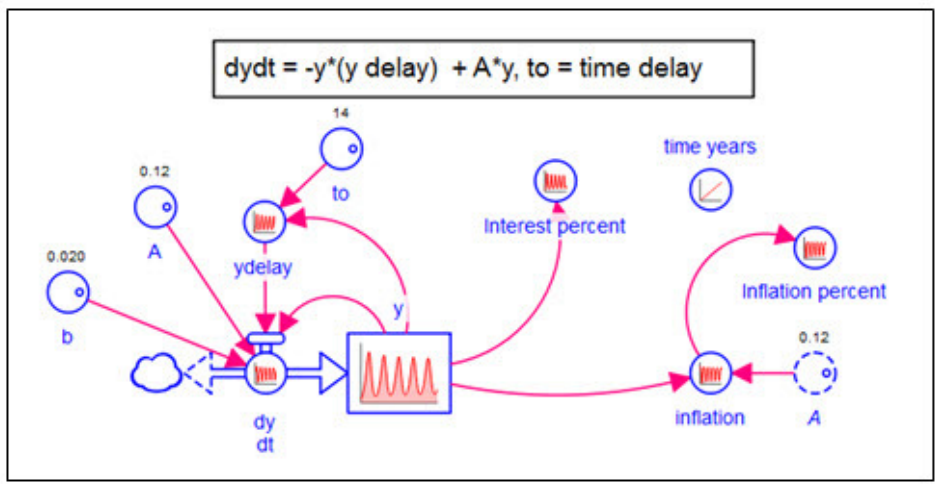

**Figure B-1.** Stella Pro Model Diagram

Here are the equations generated by Stella Pro that correspond to the variables in Figure B-1. I have added some explanations within brackets.

y(t) = y(t - dt) + (dy_dt) * dt {The variable dy/dt is represented within Stella Pro by dy_dt. The stock (box) labeled y is an integral of the inflow dy/dt}

   INIT y = b {The initial function is (b)*(A-b)*EXP(time*(A-b)). y=b at time=0.}

   INFLOWS:

     dy_dt = IF TIME >= 1 THEN -y*ydelay + y*A ELSE (b)*(A-b)*EXP(time*(A-b)) {The initial function needs to be differentiated when included in dy/dt because it will be integrated in the stock y(t). That is why we have the factor A-b in front of the exponential.}

A = 0.12

b = 0.02

inflation = A-y {Fisher's relationship}

Inflation_percent = inflation*100 {converts inflation to a percent}

Interest_percent = y*100 {converts y to a percent}

time_years = TIME/12 {converts months to years}

   to = 14 {delay in months}
   ydelay = DELAY(y,to)

## Acknowledgement

The authors would like to thank Professor Richard DeMong for providing helpful comments on our analysis.